\documentclass[twocolumn]{aastex63}

\begin{document}

\newcommand{\kms}{km~s$^{-1}$}	\newcommand{\cms}{cm~s$^{-2}$}
\newcommand{\msun}{$M_{\odot}$} \newcommand{\rsun}{$R_{\odot}$} 
\newcommand{\teff}{$T_{\rm eff}$} \newcommand{\logg}{$\log{g}$} 
\newcommand{\mas}{mas~yr$^{-1}$}


\title{The Discovery of Two LISA Sources within 0.5 kpc}

\author[0000-0001-6098-2235]{Mukremin Kilic} \affiliation{Homer L. Dodge Department 
of Physics and Astronomy, University of Oklahoma, 440 W. Brooks St., Norman, OK, 
73019 USA}

\author[0000-0002-4462-2341]{Warren R.\ Brown} \affiliation{Smithsonian 
Astrophysical Observatory, 60 Garden Street, Cambridge, MA 02138 USA}

\author[0000-0002-2384-1326]{A.\ B{\'e}dard} \affiliation{D{\'e}partement de 
Physique, Universit{\'e} de Montr{\'e}al, C.P. 6128, Succ. Centre-Ville, 
Montr{\'e}al, Quebec H3C 3J7, Canada}

\author[0000-0002-9878-1647]{Alekzander Kosakowski} \affiliation{Homer L. Dodge 
Department of Physics and Astronomy, University of Oklahoma, 440 W. Brooks St., 
Norman, OK, 73019 USA}

\email{kilic@ou.edu, wbrown@cfa.harvard.edu,\\ bedard@astro.umontreal.ca, alekzanderkos@ou.edu}
\shortauthors{Kilic, Brown, B{\'e}dard, Kosakowski}
\shorttitle{Nearby LISA Sources}

\begin{abstract}

We report the discovery of the brightest detached binary white dwarfs with periods less than an hour, which provide
two new gravitational wave verification binaries for the Laser Interferometer Space Antenna (LISA). The first one,
SMSS J033816.16$-$813929.9 (hereafter J0338), is a 30.6 min orbital period, $g=17.2$ mag detached double white dwarf
binary with a Gaia parallax measurement that places it at a distance of 533 pc. The observed radial velocity and
photometric variability provide precise constraints on the system parameters. J0338 contains a $0.230 \pm 0.015~M_{\odot}$
white dwarf with a $0.38_{-0.03}^{+0.05}~M_{\odot}$ companion at an inclination of $69 \pm 9^{\circ}$. The second system,
SDSS J063449.92+380352.2 (hereafter J0634), is a 26.5 min orbital period, $g=17.0$ mag detached double white dwarf
binary at a distance of 435 pc. J0634 contains a $0.452^{+0.070}_{-0.062}~M_{\odot}$ white dwarf with a
$0.209^{+0.034}_{-0.021}~M_{\odot}$ companion at an inclination of $37 \pm 7^{\circ}$. The more massive white
dwarf in J0634 is hotter than its companion, even though tidal dissipation is predicted to be relatively inefficient at such
periods.  This suggests that the more massive white dwarf formed last.
J0338 and J0634 will be detected by LISA with a signal-to-noise ratio of 5 and 19, respectively, after four years. 
We identified these two systems based on their overluminosity and $u$-band photometry. Follow-up of $u$-band selected
Gaia targets will likely yield additional LISA  verification binaries. 

\end{abstract}


\section{INTRODUCTION}

The Laser Interferometer Gravitational-Wave Observatory (LIGO) detected gravitational waves for the first time
in late 2015, opening a new window into the Universe \citep{ligo16}. The source was a binary black hole
merger 410 Mpc away. The Laser Interferometer Space Antenna (LISA) mission will expand this window into the mHz frequency
range, and will individually resolve tens of thousands of gravitational wave sources, including (but not limited to) ultra-compact binaries,
supermassive black hole mergers, and extreme mass ratio inspirals \citep{lisa12}. 

Ultra-compact binary white dwarfs are the dominant source class in the mHz gravitational wave frequency range
\citep{nelemans01b,nissanke12}. Based on binary population synthesis models, LISA will individually resolve
$\sim$25,000 double white dwarfs in the Galaxy \citep{korol17}. The number of double white dwarfs
that will be detected in both gravitational wave and electromagnetic observations is much smaller.
\citet{korol17} estimate 13-24 and 50-73 combined gravitational wave (with a signal-to-noise ratio of $SNR>7$)
and optical detections from LISA + Gaia and LISA + Vera Rubin Observatory's Legacy Survey of Space
and Time (LSST), respectively. 

Detection of both electromagnetic and gravitational wave signals will provide a fantastic opportunity to characterize the
double white dwarf population of the Galaxy through precise constraints on the system parameters. For example, knowing
the position and the inclination of a  binary from electromagnetic data can reduce the uncertainty in
the gravitational wave amplitude by a factor of $>40$ \citep{shah12,shah13}. A precise distance measurement,
e.g. from Gaia, can constrain the chirp mass to $\sim$15\%-25\% \citep{shah14}, while adding orbital decay measurements
($\dot{f}$) reduces it to 0.1\%. For the double white dwarf system J1539+5027 \citep{burdge19a}, \citet{littenberg19} estimate that the
combination of gravitational wave and optical measurements will improve its inclination uncertainty by a factor of 5 and
its distance uncertainty by a factor of 10.

The Extremely Low-Mass (ELM) Survey discovered the first known examples of detached white dwarf binaries
that are strong LISA sources with predicted SNR = 40-90 \citep{brown11,brown20a,kilic14}.  Recently,
\citet{burdge19a,burdge19b,burdge20a,burdge20b} discovered three additional detached double white dwarf
LISA sources with SNR $\approx$ 90 based on variable systems identified in the Zwicky Transient Facility \citep[ZTF,][]{ztf}.
Their sample includes six additional systems with predicted SNR $\sim$2-8.  
 
As part of our efforts to identify nearby LISA verification binaries, we have been following up ELM white dwarf candidates
identified through Gaia parallaxes and $ugriz$ photometry \citep{kosakowski20} from the SkyMapper Southern Survey
\citep[SMSS,][]{skymapper19} and the Sloan Digital Sky Survey \citep[SDSS,][]{ahumada20}. Here we present the discovery
of two new LISA sources within $\approx$ 0.5 kpc. We present our discovery observations
and analysis of J0338 and J0634 in Sections 2 and 3, respectively. We discuss the physical parameters of these
binaries and their expected gravitational wave signatures in Section 4, and conclude.

\begin{figure} 
\includegraphics[width=3.3in, clip, bb=10 180 570 690]{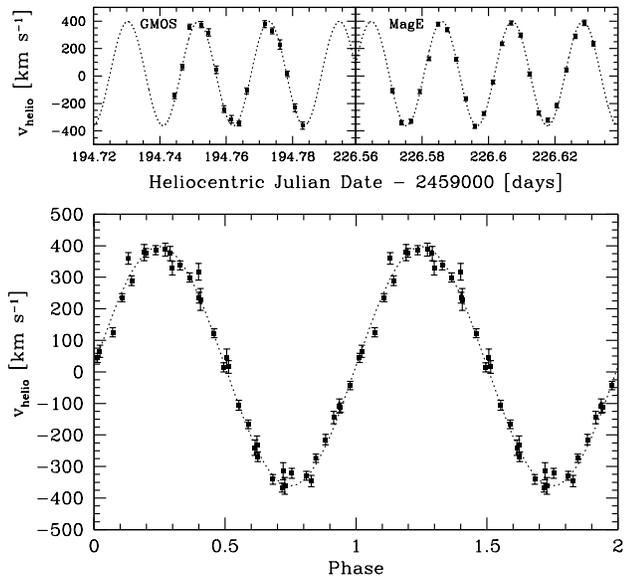}
\caption{Radial velocities of J0338 along with the best-fitting model for a circular orbit (dotted line).
The bottom panel shows all of the data points phased with the best-fit period.}
\label{figrv}
\end{figure}

\begin{figure} 
\includegraphics[width=3.25in, clip, bb=10 155 570 695]{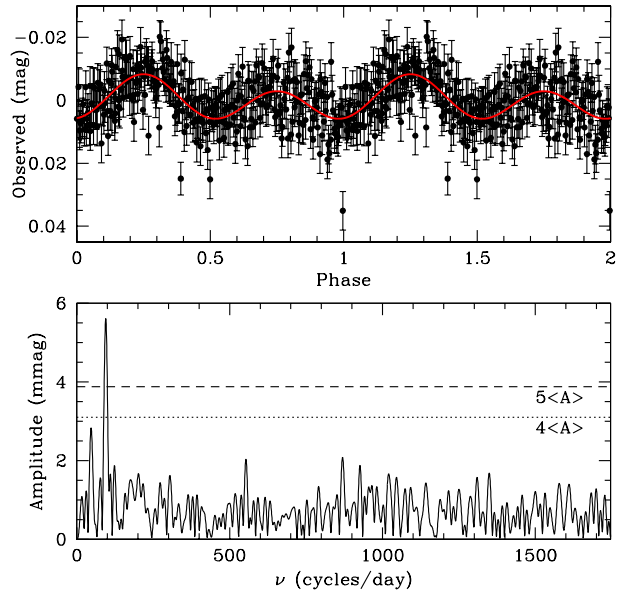}
\caption{Gemini $g$-band light curve of J0338 (top panel) folded at the orbital period, and repeated for clarity.
The solid red line displays our best-fitting model. The bottom panel shows the Fourier transform of this light curve, and
the 4$\langle {\rm A}\rangle$ and 5$\langle {\rm A}\rangle$ significance levels, where $\langle {\rm A}\rangle$ is the average
amplitude in the Fourier Transform.} 
\label{figfold}
\end{figure}

\section{SMSS J033816.16$-$813929.9}

\subsection{Radial Velocity Variability}

We obtained optical spectroscopy of J0338 using the 8m Gemini South telescope equipped with the Gemini
Multi-Object Spectrograph (GMOS) as part of the queue program GS-2020B-Q-304 on UT 2020 December 11.
We used the B600 grating and a 0.5$\arcsec$ slit, providing wavelength coverage from 3600 \AA\ to 6735 \AA\ with
a resolving power of $R=1688$. We obtained $17\times180$ s back-to-back exposures over an hour. 

Our Gemini data revealed significant radial velocity variability over a 30 min period. To constrain the
orbital period better, we obtained additional spectroscopy on UT 2021 January 12 using the MagE
instrument on the 6.5m Magellan Baade telescope. We used the 0.85 arcsec slit, providing a resolving
power of $R=4800$. We obtained $23\times215$ s back-to-back exposures over 90 min. 

Figure \ref{figrv} shows our radial velocity measurements for J0338 along with the best-fitting
circular orbit. J0338 shows 759 \kms\ peak-to-peak radial velocity variations with a 30.6 min period. 
Our model fits to its Balmer line profiles, under the assumption of a single star, indicate
$T_{\rm eff} = 18770 \pm  270$ K and $\log{g} = 6.60 \pm 0.04$,
which correspond to $M = 0.230 \pm 0.015~M_{\odot}$ and $R= 0.0396 \pm 0.0018~R_{\odot}$ based on \citet{althaus13}
models. \citet{istrate16} evolutionary sequences give results consistent with these estimates.
Given the mass function, the companion white dwarf has $M \geq 0.34 M_{\odot}$. 

\begin{figure*} 
\includegraphics[height=7.0in, angle=270, clip, bb=140 60 490 740]{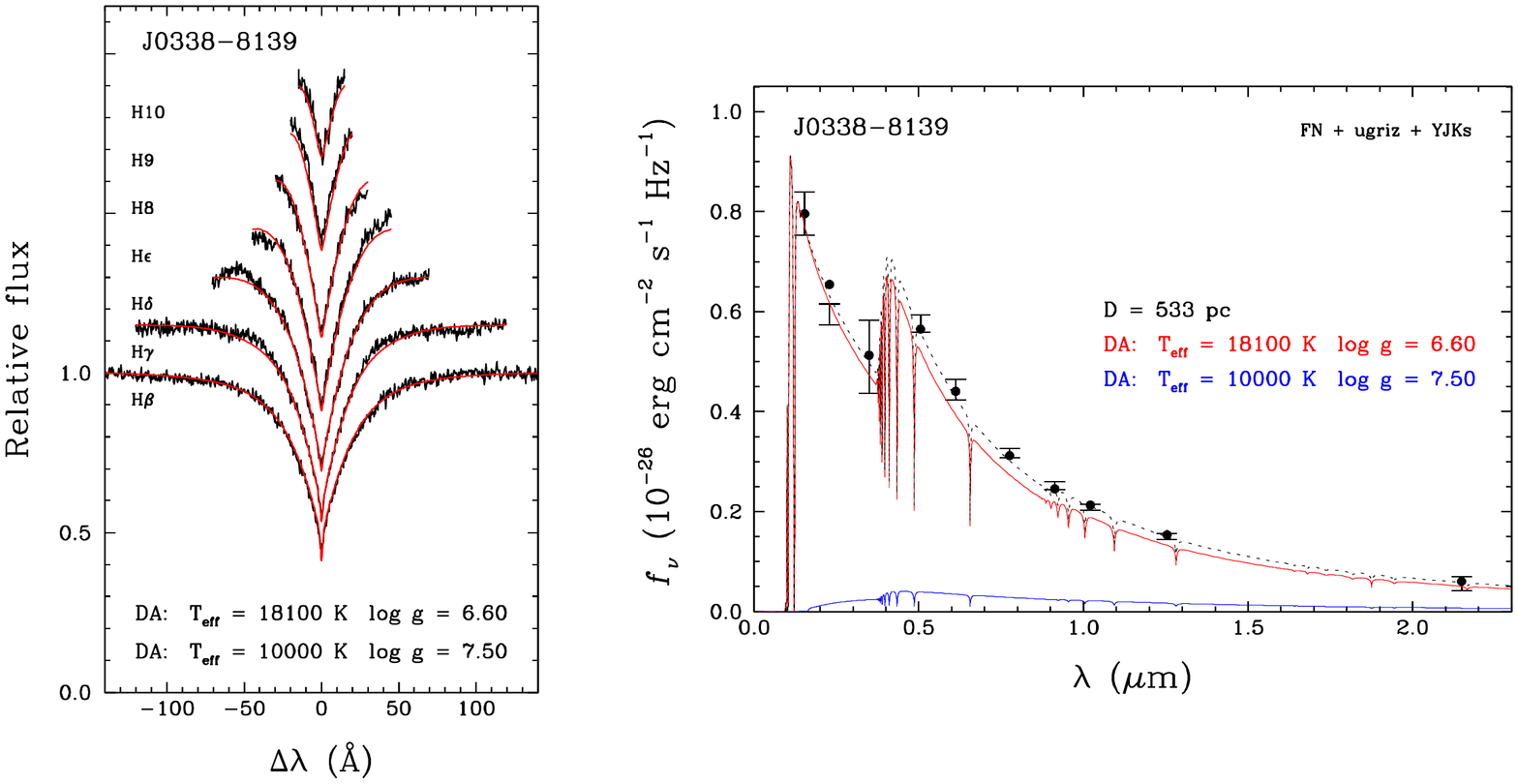}
\caption{Best joint fit to J0338's Balmer lines (left panel) and spectral energy distribution (right panel).
The left panel shows the synthetic model (red) overplotted on the observed spectrum (black).
The right panel shows the synthetic fluxes (filled circles) and observed fluxes (error bars). The
red and blue lines show the contribution of each white dwarf to the total monochromatic model flux,
displayed as the black dotted line.}
\label{figbedard}
\end{figure*}

\subsection{Ellipsoidal Variations}

We obtained Gemini GMOS time-series photometry of J0338 on UT 2021 January 25 as part of the program
GS-2021A-FT-202. We obtained $299\times7$ s back-to-back exposures through an SDSS-$g$ filter. We binned
the chip by 4$\times$4, which resulted in a plate scale of $0.32\arcsec$ pixel$^{-1}$ and a 17.7 s overhead,
resulting in a cadence of 24.7 s.

Figure \ref{figfold} shows our Gemini light curve of J0338 and its Fourier transform. The latter shows a significant peak at
$94.3 \pm 0.6$ cycles d$^{-1}$, which is half the orbital period measured from the radial velocity data. There is also
a smaller peak at the orbital frequency. The top panel shows the light curve folded at the orbital
period, and repeated for clarity. J0338 shows evidence of both the Doppler beaming effect \citep{zucker07} and
ellipsoidal variations due to a tidally distorted white dwarf.

We perform a simultaneous, non-linear least-squares fit that includes the amplitude of the Doppler beaming ($\sin{\phi}$),
ellipsoidal variations ($\cos{2\phi}$), and reflection ($\cos{\phi}$). We find the amplitudes of these effects to
be $0.25 \pm 0.05$\%, $0.52 \pm 0.05$\%, and $0_{-0.00}^{+0.03}$\%, respectively. The amplitude of the Doppler
beaming effect is mainly determined by the temperature and radial velocity semi-amplitude of the ELM white dwarf
\citep{shporer10}. The predicted Doppler beaming amplitude for J0338 is exactly the same as the observed amplitude of 0.25\%.

The ellipsoidal variation amplitude mainly depends on the mass ratio of the binary, the radius of the primary, and the
inclination \citep{morris93}. The former can be constrained based on the radial velocity variations,
and the radius of the primary can be constrained directly using the spectral energy distribution and fitting for the solid angle
$\pi (R/D)^2$, where $R$ is the radius of the star and $D$ is its distance \citep[$D=533^{+13}_{-14}$ pc for J0338,][]{bailer21}.
Hence, the amplitude of the ellipsoidal variations can be used
to constrain the inclination of the binary. For J0338, the $0.52 \pm 0.05$\% ellipsoidal variation amplitude requires
an inclination of $69 \pm 9^{\circ}$. The lack of eclipses also sets an upper limit on inclination as $\leq78^{\circ}$.
Hence, based on these constraints, the mass of the companion white dwarf is $0.38_{-0.03}^{+0.05}~M_{\odot}$.

\subsection{Binary Parameters}

Our single-star analysis of the Balmer lines of J0338 yields a spectroscopic distance in good agreement with the Gaia EDR3 parallax, indicating that the primary component largely dominates the observed flux of the system. For this reason,
the temperature, radius, and mass of the ELM white dwarf in J0338 are reliably constrained based on optical
spectroscopy and Gaia parallax. The mass of the secondary white dwarf and the inclination of the binary
are also constrained based on the radial velocity data and the observed ellipsoidal variations. 

To constrain the temperature of the secondary white dwarf, we rely on the deconvolution procedure introduced
by \citet{bedard17}, where we fit simultaneously the observed Balmer lines and spectral energy distribution with
composite model atmospheres. We use de-reddened GALEX FUV and NUV \citep{martin05}, SkyMapper $ugriz$
\citep{skymapper19}, and VISTA Hemisphere Survey $YJK_s$ photometry \citep{mcmahon21}. 
Since the observed flux can be expressed as a combination of the individual contributions of the two components, the free parameters
usually involved in this fitting process are the four atmospheric parameters $T_{\rm eff,1}, \log{g_1}, T_{\rm eff,2}, \log{g_2}$, plus
the distance $D$ \citep[see][for details]{bedard17,kilic20b}. In the present case, the distance is known from the Gaia parallax, and the
surface gravities can be obtained from our mass estimates together with the mass-radius relation of \citet{althaus13}.
Consequently, we fit only for the effective temperatures while holding other parameters fixed. Even though $T_{\rm eff,1}$ is relatively
well constrained from our single-star fit to the Balmer lines, it is allowed to vary as it is slightly affected by the addition of a secondary star.

Given our inclination constraints from ellipsoidal variations, there exists a unique fit to the Balmer lines and the spectral
energy distribution of this system. Figure \ref{figbedard} shows this best-fitting solution, where the combination of a
$18100 \pm 300$ K primary and a $10000 \pm 1000$ K companion provides an excellent match to the data. The secondary
star is $14\times$ fainter than the primary star in the $g$-band and hence barely affects the Balmer lines. However, it makes
a small contribution to the spectral energy distribution, with the result that the primary white dwarf must be slightly cooler than
indicated by the single-star analysis.

\section{SDSS J063449.92+380352.2}

\subsection{Radial Velocity and Photometric Variability}

\begin{figure} 
\includegraphics[width=3.3in, clip, bb=10 180 570 690]{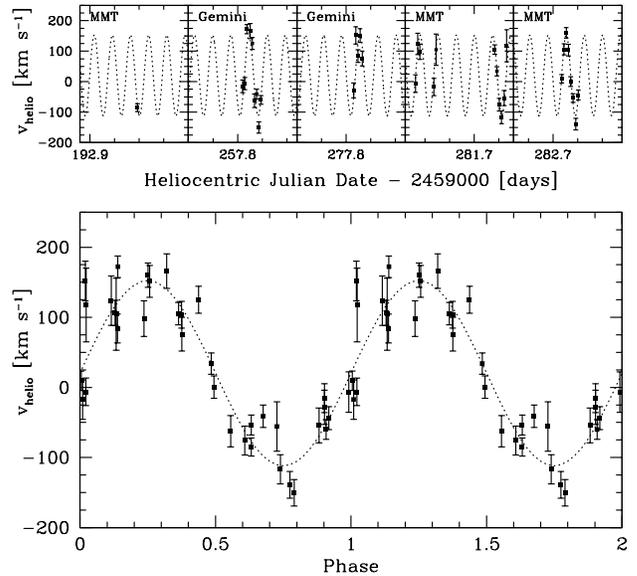}
\caption{Radial velocities of J0634 along with the best-fitting model for a circular orbit (dotted line).
The bottom panel shows all of the data points phased with the best-fit period.}
\label{figrv6}
\end{figure}

We obtained a single spectrum of J0634 using the 6.5m MMT with the Blue Channel spectrograph on UT 2020 December 9.
We operated the spectrograph with the 832 line mm$^{-1}$ grating in second order and a 1$\arcsec$ slit, providing wavelength
coverage from 3600 \AA\ to 4500 \AA\ and a spectral resolution of 1.0 \AA. This initial spectrum confirmed J0634 as
a low-mass white dwarf.

We obtained time-series optical spectroscopy of J0634 using the 8m Gemini North telescope equipped with GMOS as part of
the queue program GN-2021A-Q-300 on UT 2021 February 12 and March 4. The observing setup was identical to that of
J0338 on Gemini South. Our Gemini data revealed significant radial velocity variability over a $\approx$ 26 min period. 

To constrain the orbital period better, we obtained additional MMT spectroscopy on March 8-9.
Figure \ref{figrv6} shows our radial velocity measurements for J0634 along with the best-fitting
circular orbit. J0634 shows 264 \kms\ peak-to-peak radial velocity variations with a 26.5 min period. 

\begin{figure} 
\includegraphics[width=3.25in, clip, bb=10 155 590 695]{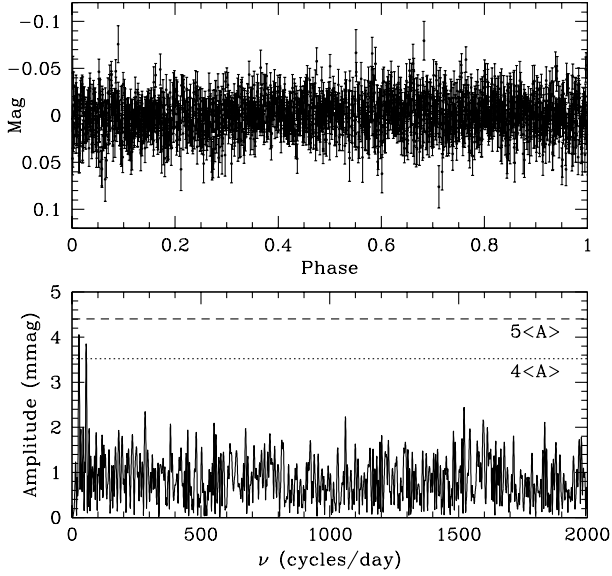}
\caption{APO 3.5m light curve of J0634 (top panel) folded at the orbital period.
The bottom panel shows the Fourier transform of this light curve, and
the 4$\langle {\rm A}\rangle$ and 5$\langle {\rm A}\rangle$ significance levels.} 
\label{figapo}
\end{figure}

We acquired high speed photometry of J0634 over 4.7 hours on UT 2021 February 26 using the APO 3.5m telescope
with the Agile frame transfer camera \citep{mukadam11} and the BG40 filter. We used an exposure time of 10 s and
binned the CCD by $2\times2$, which resulted in a plate scale of 0.258 arcsec pixel$^{-1}$. 

Figure \ref{figapo} shows our APO light curve of J0634 (top panel) and its Fourier transform (bottom panel). The latter
shows two peaks at the 4$<$A$>$ level, at $0.5\times$ and $1\times$ the orbital frequency. The peak at half the orbital frequency
is likely an artifact of the relatively short observing baseline. A Monte Carlo analysis using the {\tt Period04} package \citep{period04}
finds an amplitude of $0.34 \pm 0.12$\% for the peak at the orbital frequency. Hence, J0634
likely shows low-level sinusoidal variability at the orbital period. Given the lack of ellipsoidal variations,
the light curve does not provide any additional constraints on the component masses.

\begin{figure}
\includegraphics[width=3.8in]{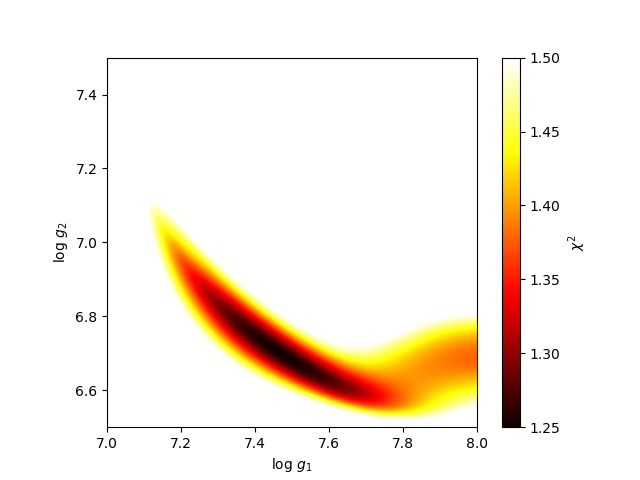}
\caption{$\chi^{2}$ distribution for our joint analysis as a function of the surface gravity of the primary and secondary star in J0634.
The dark region corresponds to the $\log{g_1}$ and $\log{g_2}$ values for which a consistent fit to the spectroscopy and photometry is possible, each solution having its own best-fitting $T_{\rm eff,1}$ and $T_{\rm eff,2}$ values.} 
\label{figchi2}
\end{figure}

\begin{figure*} 
\includegraphics[height=7.0in, angle=270, clip, bb=140 60 490 740]{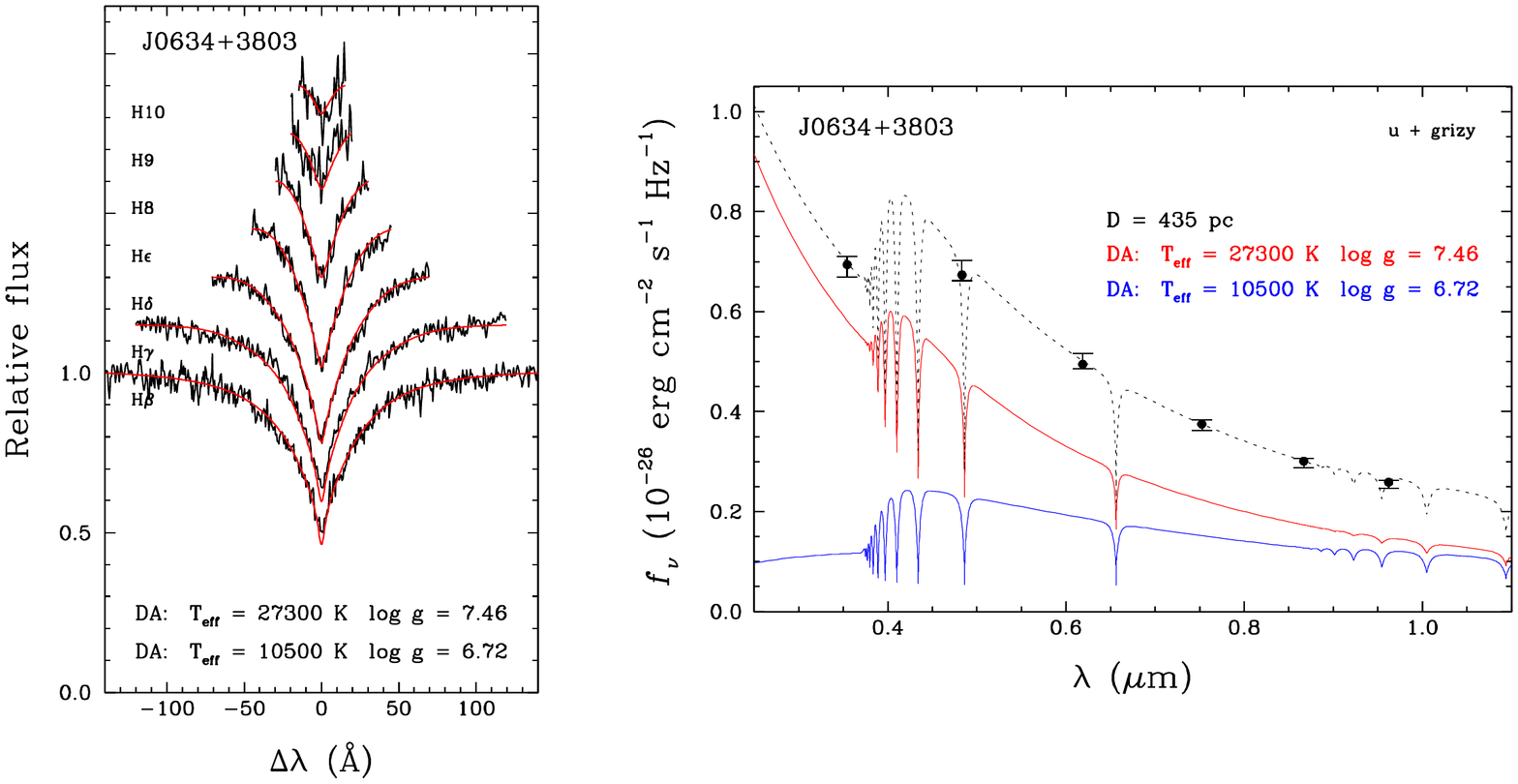}
\caption{Best joint fit to J0634's Balmer lines (left panel) and spectral energy distribution (right panel).
The left panel shows the synthetic model (red) overplotted on the observed spectrum (black).
The right panel shows the synthetic fluxes (filled circles) and observed fluxes (error bars). The
red and blue lines show the contribution of each white dwarf to the total monochromatic model flux,
displayed as the black dotted line.}
\label{fit0634}
\end{figure*}

\subsection{Binary Parameters}
\label{sec0634}

The spectroscopic distance estimate for J0634, under the assumption of a single object dominating the light, differs
significantly from the Gaia EDR3 parallax, indicating that the companion contributes significant light in this system.
Based on the deconvolution procedure from \citet{bedard17}, fitting the observed Balmer lines and de-reddened
SDSS $u$ and Pan-STARRS $grizy$ photometry with $T_{\rm eff,1}, \log{g_1}, T_{\rm eff,2}$, and $\log{g_2}$ all
allowed to vary, we find several possible solutions that reproduce the  the spectroscopic and photometric data for J0634.

Figure \ref{figchi2} shows the $\chi^2$ distribution of our fits as a function of $\log{g_1}$ and $\log{g_2}$. The dark
region corresponds to the $\log{g_1}$ and $\log{g_2}$ values for which a consistent fit to the spectroscopy and
photometry is possible, each solution having its own best-fitting $T_{\rm eff,1}$ and $T_{\rm eff,2}$ values. There are
many possible $\log{g_1}$ and $\log{g_2}$ solutions along a narrow banana shaped region. The brighter (primary) star
in the system has a higher surface gravity compared to the companion. However, an increase in $\log{g_1}$ can be
compensated by a decrease in $\log{g_2}$. In other words, a more massive, smaller, and fainter primary requires a less
massive, larger, and brighter companion white dwarf to explain the observed spectral energy distribution.

Figure \ref{fit0634} shows the best-fitting solution that lies at the center of the minimum $\chi^2$ region in Figure \ref{figchi2}.
The observed Balmer lines and $ugrizy$ photometry are explained fairly well by a binary system of a primary
white dwarf with $(T_{\rm eff}, \log{g}, M)_1$ = (27300 K, 7.46, 0.452 $M_{\odot})$ and a secondary white
dwarf with $(T_{\rm eff}, \log{g}, M)_2$ = (10500 K, 6.72, 0.209 $M_{\odot}$), where the mass estimates
are based on the He-core models of \citet{althaus13}. For comparison, the best-fitting solution at the top left boundary
of the dark region in Figure \ref{figchi2} has $(T_{\rm eff}, \log{g}, M)_1$ = (24400 K, 7.24, 0.390 $M_{\odot}$), and
$(T_{\rm eff}, \log{g}, M)_2$ = (10300 K, 6.91, 0.243 $M_{\odot}$), whereas the best-fitting solution at the bottom right
boundary has $(T_{\rm eff}, \log{g}, M)_1$ = (31300 K, 7.74, 0.522 $M_{\odot}$), and 
$(T_{\rm eff}, \log{g}, M)_2$ = (10800 K, 6.59, 0.188 $M_{\odot}$).
In the latter case, the primary lies outside of the parameter space covered by the \citet{althaus13} cooling tracks,
so we instead used the CO-core models of \citet{bedard20} to evaluate $M_1$.

As expected, for all viable solutions, the secondary star contributes appreciably to the total flux. We verified that
the best-fitting solution does not predict significant ellipsoidal variations ($\leq0.01$\%) and a detectable double H$\alpha$
feature at the relatively low resolution of our Gemini spectra. The best-fitting solution would lead to 0.07\% and
$\sim$0.35\% amplitude \citep{morris93} relativistic beaming and reflection effect, respectively. The latter is consistent
with the $0.34 \pm 0.12$\% observed variability in the APO light curve of J0634. However, additional time series photometry
is needed to confirm the source of variability in this system. We conclude that the primary
and secondary white dwarfs in J0634 have masses of $0.452^{+0.070}_{-0.062}~M_{\odot}$ and
$0.209^{+0.034}_{-0.021}~M_{\odot}$, respectively. Based on the mass function, the orbital inclination
of this system is $i=37 \pm 7^{\circ}$.

\section{Discussion}

\begin{figure}
 \includegraphics[width=3.3in, clip, bb=10 135 570 700]{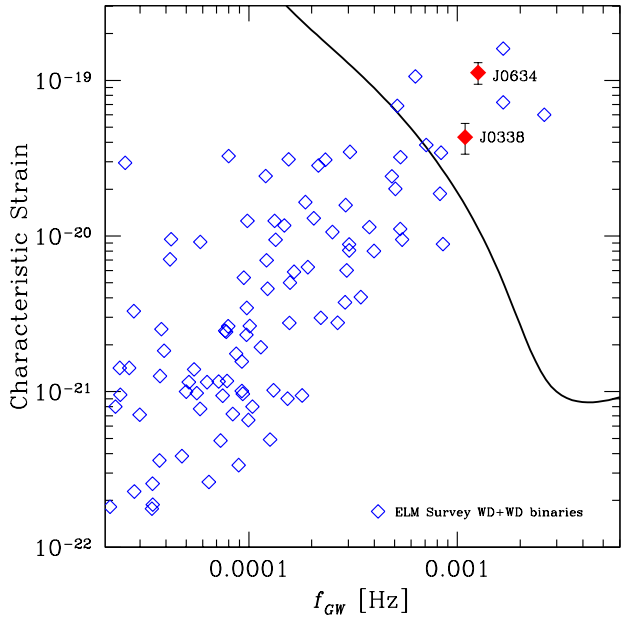}
 \caption{ \label{figgw}
Characteristic strain versus gravitational wave frequency for J0338, J0634, and other detached 
white dwarf binaries found in the ELM Survey \citep[open diamonds,][]{brown20b}.
Solid line is the LISA 4~yr sensitivity curve \citep{robson19}.}
 \end{figure}

J0338 and J0634 are newly identified ultra-compact binary systems within $\approx$ 0.5 kpc. Both systems involve
an ELM white dwarf with a $\sim$ 0.4 \msun\ white dwarf companion. Figure \ref{figgw} shows their characteristic
strain relative to the 4-year LISA sensitivity curve \citep{robson19}. Open diamonds are previously identified detached
binary white dwarfs found in the ELM Survey \citep{brown20b}.  According to the LISA Detectability 
Calculator\footnote{\url{https://heasarc.gsfc.nasa.gov/lisa/lisatool/}}, J0338 and J0634 are estimated to have a 4-yr 
SNR = 5 and 19, respectively. 

\citet{brown20a} identified the 20 min period binary J2322+0509 as the first He + He white dwarf system
among the LISA verification binaries. \citet{burdge20a,burdge20b} identified several additional He + He
white dwarf LISA sources. J0338 and J0634 join this list as the brightest and closest members. In fact, J0338
and J0634 are the brightest detached binary white dwarfs currently known with a period less than an hour.

Since J0338 shows ellipsoidal variations, follow-up time-series photometry over several years can constrain the
rate of period change \citep[e.g.,][]{hermes12,burdge19b}. With a $\dot{P}$ measurement and the already
available constraints on the component masses, orbital inclination, and distance, future gravitational wave
measurements will open the door for measuring tidal dissipation in white dwarfs
\citep{fuller13,piro19}. 

\begin{deluxetable}{lcc}
\tablecolumns{3}
\tablewidth{0pt}
\tablecaption{System parameters.\label{tab:param}}
\tablehead{ \colhead{Parameter} & \colhead{Value} & \colhead{Value} }
\startdata
Name & J0338 & J0634 \\
RA & 03:38:16.16 & 06:34:49.92\\
DEC & $-$81:39:29.9 & +38:03:52.2\\
$d$ (pc)		& $533^{+13}_{-14}$  & $435^{+17}_{-15}$\\
$u$ (mag)		& 17.426 $\pm$ 0.144 & 17.111 $\pm$ 0.018 \\
$g$ (mag)		& 17.204 $\pm$ 0.009 & 17.001 $\pm$ 0.010 \\
$r$ (mag)		& 17.438 $\pm$ 0.047 & 17.285 $\pm$ 0.009 \\
$E(B-V)$ (mag)		& 0.059 & 0.153 \\ 
$P$ (s)              & 1836.1  $\pm$  31.9 & 1591.4 $\pm$ 28.9 \\
$K$ (\kms)	&  379.7 $\pm$ 4.6 & 132.1 $\pm$ 6.0\\
$\gamma$ (\kms) & 18.0 $\pm$ 3.8   & 20.3 $\pm$ 4.3 \\
$T_{\rm eff,1}$ (K)	&  $18100 \pm 300$ & $27300^{+4000}_{-2900}$ \\
$\log{g_1}$ (cm s$^{-2}$) & $6.60 \pm 0.04$ & $7.46^{+0.28}_{-0.22}$ \\
$M_{\rm 1}$ (\msun) & $0.230 \pm 0.015$ & $0.452^{+0.070}_{-0.062}$\\
$T_{\rm eff,2}$ (K)	&  $10000 \pm 1000$ & $10500^{+300}_{-200}$ \\
$\log{g_2}$ (cm s$^{-2}$) & $7.50^{+0.12}_{-0.09}$ &  $6.72^{+0.19}_{-0.13}$\\
$M_{\rm 2}$ (\msun) & $0.38_{-0.03}^{+0.05}$ & $0.209^{+0.034}_{-0.021}$ \\
$i$ ($^{\circ}$)		& $69 \pm 9$ & $37 \pm 7$ \\
\enddata
\tablenote{Note: The primary and secondary star parameters in J0634 are highly anti-correlated (see Figure \ref{figchi2}
and Section \ref{sec0634}).}
\end{deluxetable}

Even though J0338 and J0634 have comparable orbital periods and component masses, the temperatures of the primary
and secondary stars are inverted in J0634, where the more massive star is significantly hotter than its companion.
Tidal energy dissipation within a white dwarf can increase its surface temperature. \citet{burdge19a} estimate that
tidal heating alone can heat up the primary white dwarf in the 7 min orbital period system, ZTF J1539+5027,
to $\approx$ 19000 K. Though, this is still much lower than the observed temperature of
48900 K, and intermittent accretion may be responsible for the observed heating of
the primary star in that system. 

\citet[][see their Fig. 8]{fuller13} show that tidal heating makes a significant contribution
to the luminosity for $\sim$ 10 min orbital period systems, but it becomes relatively insignificant at
$\sim$ 30 min periods since $T_{\rm tide} \propto (\frac{|\dot{P}|}{P^3})^{1/4}$. Hence, tidal heating cannot
explain the inverted temperatures of the primary and secondary stars in J0634. 

There are other examples of short period double white dwarfs with inverted temperatures between the primary
and secondary stars \citep{bours15}. For example, ZTF J1749+0924 is similar to J0634 in its characteristics;
it has an orbital period of 26.43 min and the 0.40 $M_{\odot}$ primary is significantly hotter than the 0.28 $M_{\odot}$
secondary \citep{burdge20a}. 

The temperature difference suggests that the more massive white dwarf formed last.
This is expected for some systems evolving through stable and then unstable mass transfer. Studying the formation
channels for double white dwarfs, \citet{ruiter10} note that the detached systems typically evolve from progenitors with
comparable masses, and that the number of detached He-CO
white dwarf binaries in which the He white dwarf forms first (i.e., the HeCO-D1 systems in their Table 1) is comparable
to the number of binaries in which the CO white dwarf forms first (the COHe-D1 systems). Hence, J0338 and J0634
may simply represent the different outcomes of double white dwarf formation channels.

There are currently 14 detached double white dwarfs known with $P<40$ min, including J0338 and J0634.
Five were discovered in the ELM Survey \citep{brown20b} and seven in the ZTF \citep{burdge20a,burdge20b}.
All 14 of these systems contain a white dwarf hotter than 16,000 K. This is not surprising. Given the $\sim$Myr
merger timescales for these ultra-compact binaries, such systems are rare, and we are more likely to find the
brightest and hottest (and therefore youngest) systems in magnitude-limited surveys. J0338 and J0634 have
$M_g =$ 8.2 - 8.4 mag and de-reddened $u-g = 0.13$ and $-0.06$ mag, respectively.  Hence, an excellent way to find
more of these systems is through follow-up of nearby blue over-luminous objects like J0338 and J0634.
 
The ELM Survey \citep{kilic10a,brown10,brown20b} and ZTF \citep{burdge19a,burdge20a,burdge20b} have explored
ELM white dwarf candidates in the northern sky. The $ugriz$ photometry from the SkyMapper Southern
Survey provides an excellent opportunity to find the brightest ELM white dwarfs in the southern sky, and we are currently
following up additional SkyMapper targets as part of the ELM Survey South \citep{kosakowski20}. The ten-year Rubin
Observatory Legacy Survey of Space and Time (LSST) and the BlackGEM \citep{groot19} survey will provide both $u$-band photometry
and multi-epoch photometry in several filters, which will provide a fantastic opportunity to significantly increase
the number of double white dwarfs that are LISA sources. 

\acknowledgements

This work was supported in part by the NSF under grant AST-1906379, the Smithsonian Institution,
the NSERC Canada, and by the Fund FRQ-NT (Qu\'ebec).

Based on observations obtained at the Gemini, MMT, and Apache Point Observatory.
Gemini is operated by the Association of Universities for Research in Astronomy, Inc., under a cooperative 
agreement with the NSF on behalf of the Gemini partnership: the National Science 
Foundation (United States), National Research Council (Canada), CONICYT (Chile), 
Ministerio de Ciencia, Tecnolog\'{i}a e Innovaci\'{o}n Productiva (Argentina), 
Minist\'{e}rio da Ci\^{e}ncia, Tecnologia e Inova\c{c}\~{a}o (Brazil), and Korea 
Astronomy and Space Science Institute (Republic of Korea).

The MMT is a joint facility of the Smithsonian Institution and the University of Arizona.
The Apache Point Observatory 3.5-meter telescope is owned and operated by the
Astrophysical Research Consortium.

\facilities{Gemini:North and South (GMOS spectrograph), Magellan:Clay (MagE spectrograph), APO 3.5m (Agile)}


\begin{thebibliography}{}
\expandafter\ifx\csname natexlab\endcsname\relax\def\natexlab#1{#1}\fi

\bibitem[{{Abbott} {et~al.}(2016){Abbott}, {Abbott}, {Abbott}, {Abernathy},
  {Acernese}, {Ackley}, {Adams}, {Adams}, {Addesso}, {Adhikari}, {Adya},
  {Affeldt}, {Agathos}, {Agatsuma}, {Aggarwal}, {Aguiar}, {Aiello}, {Ain},
  {Ajith}, {Allen}, {Allocca}, {Altin}, {Anderson}, {Anderson}, {Arai},
  {Arain}, {Araya}, {Arceneaux}, {Areeda}, {Arnaud}, {Arun}, {Ascenzi},
  {Ashton}, {Ast}, {Aston}, {Astone}, {Aufmuth}, {Aulbert}, {Babak}, {Bacon},
  {Bader}, {Baker}, {Baldaccini}, {Ballardin}, {Ballmer}, {Barayoga},
  {Barclay}, {Barish}, {Barker}, {Barone}, {Barr}, {Barsotti}, {Barsuglia},
  {Barta}, {Bartlett}, {Barton}, {Bartos}, {Bassiri}, {Basti}, {Batch},
  {Baune}, {Bavigadda}, {Bazzan}, {Behnke}, {Bejger}, {Belczynski}, {Bell},
  {Bell}, {Berger}, {Bergman}, {Bergmann}, {Berry}, {Bersanetti}, {Bertolini},
  {Betzwieser}, {Bhagwat}, {Bhandare}, {Bilenko}, {Billingsley}, {Birch},
  {Birney}, {Birnholtz}, {Biscans}, {Bisht}, {Bitossi}, {Biwer}, {Bizouard},
  {Blackburn}, {Blair}, {Blair}, {Blair}, {Bloemen}, {Bock}, {Bodiya}, {Boer},
  {Bogaert}, {Bogan}, {Bohe}, {Bojtos}, {Bond}, {Bondu}, {Bonnand}, {Boom},
  {Bork}, {Boschi}, {Bose}, {Bouffanais}, {Bozzi}, {Bradaschia}, {Brady},
  {Braginsky}, {Branchesi}, {Brau}, {Briant}, {Brillet}, {Brinkmann},
  {Brisson}, {Brockill}, {Brooks}, {Brown}, {Brown}, {Brown}, {Buchanan},
  {Buikema}, {Bulik}, {Bulten}, {Buonanno}, {Buskulic}, {Buy}, {Byer},
  {Cabero}, {Cadonati}, {Cagnoli}, {Cahillane}, {Bustillo}, {Callister},
  {Calloni}, {Camp}, {Cannon}, {Cao}, {Capano}, {Capocasa}, {Carbognani},
  {Caride}, {Casanueva Diaz}, {Casentini}, {Caudill}, {Cavagli{\`a}},
  {Cavalier}, {Cavalieri}, {Cella}, {Cepeda}, {Baiardi}, {Cerretani},
  {Cesarini}, {Chakraborty}, {Chalermsongsak}, {Chamberlin}, {Chan}, {Chao},
  {Charlton}, {Chassande-Mottin}, {Chen}, {Chen}, {Cheng}, {Chincarini},
  {Chiummo}, {Cho}, {Cho}, {Chow}, {Christensen}, {Chu}, {Chua}, {Chung},
  {Ciani}, {Clara}, {Clark}, {Cleva}, {Coccia}, {Cohadon}, {Colla}, {Collette},
  {Cominsky}, {Constancio}, {Conte}, {Conti}, {Cook}, {Corbitt}, {Cornish},
  {Corsi}, {Cortese}, {Costa}, {Coughlin}, {Coughlin}, {Coulon}, {Countryman},
  {Couvares}, {Cowan}, {Coward}, {Cowart}, {Coyne}, {Coyne}, {Craig},
  {Creighton}, {Creighton}, {Cripe}, {Crowder}, {Cruise}, {Cumming},
  {Cunningham}, {Cuoco}, {Dal Canton}, {Danilishin}, {D'Antonio}, {Danzmann},
  {Darman}, {Da Silva Costa}, {Dattilo}, {Dave}, {Daveloza}, {Davier},
  {Davies}, {Daw}, {Day}, {De}, {DeBra}, {Debreczeni}, {Degallaix}, {De
  Laurentis}, {Del{\'e}glise}, {Del Pozzo}, {Denker}, {Dent}, {Dereli},
  {Dergachev}, {DeRosa}, {De Rosa}, {DeSalvo}, {Dhurandhar}, {D{\'\i}az}, {Di
  Fiore}, {Di Giovanni}, {Di Lieto}, {Di Pace}, {Di Palma}, {Di Virgilio},
  {Dojcinoski}, {Dolique}, {Donovan}, {Dooley}, {Doravari}, {Douglas},
  {Downes}, {Drago}, {Drever}, {Driggers}, {Du}, {Ducrot}, {Dwyer}, {Edo},
  {Edwards}, {Effler}, {Eggenstein}, {Ehrens}, {Eichholz}, {Eikenberry},
  {Engels}, {Essick}, {Etzel}, {Evans}, {Evans}, {Everett}, {Factourovich},
  {Fafone}, {Fair}, {Fairhurst}, {Fan}, {Fang}, {Farinon}, {Farr}, {Farr},
  {Favata}, {Fays}, {Fehrmann}, {Fejer}, {Feldbaum}, {Ferrante}, {Ferreira},
  {Ferrini}, {Fidecaro}, {Finn}, {Fiori}, {Fiorucci}, {Fisher}, {Flaminio},
  {Fletcher}, {Fong}, {Fournier}, {Franco}, {Frasca}, {Frasconi}, {Frede},
  {Frei}, {Freise}, {Frey}, {Frey}, {Fricke}, {Fritschel}, {Frolov}, {Fulda},
  {Fyffe}, {Gabbard}, {Gair}, {Gammaitoni}, {Gaonkar}, {Garufi}, {Gatto},
  {Gaur}, {Gehrels}, {Gemme}, {Gendre}, {Genin}, {Gennai}, {George}, {Gergely},
  {Germain}, {Ghosh}, {Ghosh}, {Ghosh}, {Giaime}, {Giardina}, {Giazotto},
  {Gill}, {Glaefke}, {Gleason}, {Goetz}, {Goetz}, {Gondan}, {Gonz{\'a}lez},
  {Castro}, {Gopakumar}, {Gordon}, {Gorodetsky}, {Gossan}, {Gosselin},
  {Gouaty}, {Graef}, {Graff}, {Granata}, {Grant}, {Gras}, {Gray}, {Greco},
  {Green}, {Greenhalgh}, {Groot}, {Grote}, {Grunewald}, {Guidi}, {Guo},
  {Gupta}, {Gupta}, {Gushwa}, {Gustafson}, {Gustafson}, {Hacker}, {Hall},
  {Hall}, {Hammond}, {Haney}, {Hanke}, {Hanks}, {Hanna}, {Hannam}, {Hanson},
  {Hardwick}, {Harms}, {Harry}, {Harry}, {Hart}, {Hartman}, {Haster},
  {Haughian}, {Healy}, {Heefner}, {Heidmann}, {Heintze}, {Heinzel}, {Heitmann},
  {Hello}, {Hemming}, {Hendry}, {Heng}, {Hennig}, {Heptonstall}, {Heurs},
  {Hild}, {Hoak}, {Hodge}, {Hofman}, {Hollitt}, {Holt}, {Holz}, {Hopkins},
  {Hosken}, {Hough}, {Houston}, {Howell}, {Hu}, {Huang}, {Huerta}, {Huet},
  {Hughey}, {Husa}, {Huttner}, {Huynh-Dinh}, {Idrisy}, {Indik}, {Ingram},
  {Inta}, {Isa}, {Isac}, {Isi}, {Islas}, {Isogai}, {Iyer}, {Izumi}, {Jacobson},
  {Jacqmin}, {Jang}, {Jani}, {Jaranowski}, {Jawahar}, {Jim{\'e}nez-Forteza},
  {Johnson}, {Johnson-McDaniel}, {Jones}, {Jones}, {Jonker}, {Ju}, {Haris},
  {Kalaghatgi}, {Kalogera}, {Kandhasamy}, {Kang}, {Kanner}, {Karki},
  {Kasprzack}, {Katsavounidis}, {Katzman}, {Kaufer}, {Kaur}, {Kawabe},
  {Kawazoe}, {K{\'e}f{\'e}lian}, {Kehl}, {Keitel}, {Kelley}, {Kells},
  {Kennedy}, {Keppel}, {Key}, {Khalaidovski}, {Khalili}, {Khan}, {Khan},
  {Khan}, {Khazanov}, {Kijbunchoo}, {Kim}, {Kim}, {Kim}, {Kim}, {Kim}, {Kim},
  {King}, {King}, {Kinzel}, {Kissel}, {Kleybolte}, {Klimenko}, {Koehlenbeck},
  {Kokeyama}, {Koley}, {Kondrashov}, {Kontos}, {Koranda}, {Korobko}, {Korth},
  {Kowalska}, {Kozak}, {Kringel}, {Krishnan}, {Kr{\'o}lak}, {Krueger}, {Kuehn},
  {Kumar}, {Kumar}, {Kuo}, {Kutynia}, {Kwee}, {Lackey}, {Landry}, {Lange},
  {Lantz}, {Lasky}, {Lazzarini}, {Lazzaro}, {Leaci}, {Leavey}, {Lebigot},
  {Lee}, {Lee}, {Lee}, {Lee}, {Lenon}, {Leonardi}, {Leong}, {Leroy},
  {Letendre}, {Levin}, {Levine}, {Li}, {Libson}, {Littenberg}, {Lockerbie},
  {Logue}, {Lombardi}, {London}, {Lord}, {Lorenzini}, {Loriette}, {Lormand},
  {Losurdo}, {Lough}, {Lousto}, {Lovelace}, {L{\"u}ck}, {Lundgren}, {Luo},
  {Lynch}, {Ma}, {MacDonald}, {Machenschalk}, {MacInnis}, {Macleod},
  {Maga{\~n}a-Sandoval}, {Magee}, {Mageswaran}, {Majorana}, {Maksimovic},
  {Malvezzi}, {Man}, {Mandel}, {Mandic}, {Mangano}, {Mansell}, {Manske},
  {Mantovani}, {Marchesoni}, {Marion}, {M{\'a}rka}, {M{\'a}rka}, {Markosyan},
  {Maros}, {Martelli}, {Martellini}, {Martin}, {Martin}, {Martynov}, {Marx},
  {Mason}, {Masserot}, {Massinger}, {Masso-Reid}, {Matichard}, {Matone},
  {Mavalvala}, {Mazumder}, {Mazzolo}, {McCarthy}, {McClelland}, {McCormick},
  {McGuire}, {McIntyre}, {McIver}, {McManus}, {McWilliams}, {Meacher},
  {Meadors}, {Meidam}, {Melatos}, {Mendell}, {Mendoza-Gandara}, {Mercer},
  {Merilh}, {Merzougui}, {Meshkov}, {Messenger}, {Messick}, {Meyers},
  {Mezzani}, {Miao}, {Michel}, {Middleton}, {Mikhailov}, {Milano}, {Miller},
  {Millhouse}, {Minenkov}, {Ming}, {Mirshekari}, {Mishra}, {Mitra},
  {Mitrofanov}, {Mitselmakher}, {Mittleman}, {Moggi}, {Mohan}, {Mohapatra},
  {Montani}, {Moore}, {Moore}, {Moraru}, {Moreno}, {Morriss}, {Mossavi},
  {Mours}, {Mow-Lowry}, {Mueller}, {Mueller}, {Muir}, {Mukherjee}, {Mukherjee},
  {Mukherjee}, {Mukund}, {Mullavey}, {Munch}, {Murphy}, {Murray}, {Mytidis},
  {Nardecchia}, {Naticchioni}, {Nayak}, {Necula}, {Nedkova}, {Nelemans},
  {Neri}, {Neunzert}, {Newton}, {Nguyen}, {Nielsen}, {Nissanke}, {Nitz},
  {Nocera}, {Nolting}, {Normandin}, {Nuttall}, {Oberling}, {Ochsner}, {O'Dell},
  {Oelker}, {Ogin}, {Oh}, {Oh}, {Ohme}, {Oliver}, {Oppermann}, {Oram},
  {O'Reilly}, {O'Shaughnessy}, {Ott}, {Ottaway}, {Ottens}, {Overmier}, {Owen},
  {Pai}, {Pai}, {Palamos}, {Palashov}, {Palomba}, {Pal-Singh}, {Pan}, {Pan},
  {Pankow}, {Pannarale}, {Pant}, {Paoletti}, {Paoli}, {Papa}, {Paris},
  {Parker}, {Pascucci}, {Pasqualetti}, {Passaquieti}, {Passuello},
  {Patricelli}, {Patrick}, {Pearlstone}, {Pedraza}, {Pedurand}, {Pekowsky},
  {Pele}, {Penn}, {Perreca}, {Pfeiffer}, {Phelps}, {Piccinni}, {Pichot},
  {Pickenpack}, {Piergiovanni}, {Pierro}, {Pillant}, {Pinard}, {Pinto},
  {Pitkin}, {Poeld}, {Poggiani}, {Popolizio}, {Post}, {Powell}, {Prasad},
  {Predoi}, {Premachandra}, {Prestegard}, {Price}, {Prijatelj}, {Principe},
  {Privitera}, {Prix}, {Prodi}, {Prokhorov}, {Puncken}, {Punturo}, {Puppo},
  {P{\"u}rrer}, {Qi}, {Qin}, {Quetschke}, {Quintero}, {Quitzow-James}, {Raab},
  {Rabeling}, {Radkins}, {Raffai}, {Raja}, {Rakhmanov}, {Ramet}, {Rapagnani},
  {Raymond}, {Razzano}, {Re}, {Read}, {Reed}, {Regimbau}, {Rei}, {Reid},
  {Reitze}, {Rew}, {Reyes}, {Ricci}, {Riles}, {Robertson}, {Robie}, {Robinet},
  {Rocchi}, {Rolland}, {Rollins}, {Roma}, {Romano}, {Romano}, {Romanov},
  {Romie}, {Rosi{\'n}ska}, {Rowan}, {R{\"u}diger}, {Ruggi}, {Ryan}, {Sachdev},
  {Sadecki}, {Sadeghian}, {Salconi}, {Saleem}, {Salemi}, {Samajdar}, {Sammut},
  {Sampson}, {Sanchez}, {Sandberg}, {Sandeen}, {Sanders}, {Sanders},
  {Sassolas}, {Sathyaprakash}, {Saulson}, {Sauter}, {Savage}, {Sawadsky},
  {Schale}, {Schilling}, {Schmidt}, {Schmidt}, {Schnabel}, {Schofield},
  {Sch{\"o}nbeck}, {Schreiber}, {Schuette}, {Schutz}, {Scott}, {Scott},
  {Sellers}, {Sengupta}, {Sentenac}, {Sequino}, {Sergeev}, {Serna},
  {Setyawati}, {Sevigny}, {Shaddock}, {Shaffer}, {Shah}, {Shahriar}, {Shaltev},
  {Shao}, {Shapiro}, {Shawhan}, {Sheperd}, {Shoemaker}, {Shoemaker}, {Siellez},
  {Siemens}, {Sigg}, {Silva}, {Simakov}, {Singer}, {Singer}, {Singh}, {Singh},
  {Singhal}, {Sintes}, {Slagmolen}, {Smith}, {Smith}, {Smith}, {Smith}, {Son},
  {Sorazu}, {Sorrentino}, {Souradeep}, {Srivastava}, {Staley}, {Steinke},
  {Steinlechner}, {Steinlechner}, {Steinmeyer}, {Stephens}, {Stevenson},
  {Stone}, {Strain}, {Straniero}, {Stratta}, {Strauss}, {Strigin}, {Sturani},
  {Stuver}, {Summerscales}, {Sun}, {Sutton}, {Swinkels}, {Szczepa{\'n}czyk},
  {Tacca}, {Talukder}, {Tanner}, {T{\'a}pai}, {Tarabrin}, {Taracchini},
  {Taylor}, {Theeg}, {Thirugnanasambandam}, {Thomas}, {Thomas}, {Thomas},
  {Thorne}, {Thorne}, {Thrane}, {Tiwari}, {Tiwari}, {Tokmakov}, {Tomlinson},
  {Tonelli}, {Torres}, {Torrie}, {T{\"o}yr{\"a}}, {Travasso}, {Traylor},
  {Trifir{\`o}}, {Tringali}, {Trozzo}, {Tse}, {Turconi}, {Tuyenbayev},
  {Ugolini}, {Unnikrishnan}, {Urban}, {Usman}, {Vahlbruch}, {Vajente},
  {Valdes}, {Vallisneri}, {van Bakel}, {van Beuzekom}, {van den Brand}, {Van
  Den Broeck}, {Vander-Hyde}, {van der Schaaf}, {van Heijningen}, {van Veggel},
  {Vardaro}, {Vass}, {Vas{\'u}th}, {Vaulin}, {Vecchio}, {Vedovato}, {Veitch},
  {Veitch}, {Venkateswara}, {Verkindt}, {Vetrano}, {Vicer{\'e}}, {Vinciguerra},
  {Vine}, {Vinet}, {Vitale}, {Vo}, {Vocca}, {Vorvick}, {Voss}, {Vousden},
  {Vyatchanin}, {Wade}, {Wade}, {Wade}, {Waldman}, {Walker}, {Wallace},
  {Walsh}, {Wang}, {Wang}, {Wang}, {Wang}, {Wang}, {Ward}, {Ward}, {Warner},
  {Was}, {Weaver}, {Wei}, {Weinert}, {Weinstein}, {Weiss}, {Welborn}, {Wen},
  {We{\ss}els}, {Westphal}, {Wette}, {Whelan}, {Whitcomb}, {White}, {Whiting},
  {Wiesner}, {Wilkinson}, {Willems}, {Williams}, {Williams}, {Williamson},
  {Willis}, {Willke}, {Wimmer}, {Winkelmann}, {Winkler}, {Wipf}, {Wiseman},
  {Wittel}, {Woan}, {Worden}, {Wright}, {Wu}, {Yablon}, {Yakushin}, {Yam},
  {Yamamoto}, {Yancey}, {Yap}, {Yu}, {Yvert}, {Zadro{\.Z}ny}, {Zangrando},
  {Zanolin}, {Zendri}, {Zevin}, {Zhang}, {Zhang}, {Zhang}, {Zhang}, {Zhao},
  {Zhou}, {Zhou}, {Zhu}, {Zucker}, {Zuraw}, {Zweizig}, {LIGO Scientific
  Collaboration}, \& {Virgo Collaboration}}]{ligo16}
{Abbott}, B.~P., {Abbott}, R., {Abbott}, T.~D., {et~al.} 2016, \prl, 116,
  061102

\bibitem[{{Ahumada} {et~al.}(2020){Ahumada}, {Allende Prieto}, {Almeida},
  {Anders}, {Anderson}, {Andrews}, {Anguiano}, {Arcodia}, {Armengaud},
  {Aubert}, {Avila}, {Avila-Reese}, {Badenes}, {Balland}, {Barger},
  {Barrera-Ballesteros}, {Basu}, {Bautista}, {Beaton}, {Beers}, {Benavides},
  {Bender}, {Bernardi}, {Bershady}, {Beutler}, {Bidin}, {Bird}, {Bizyaev},
  {Blanc}, {Blanton}, {Boquien}, {Borissova}, {Bovy}, {Brandt}, {Brinkmann},
  {Brownstein}, {Bundy}, {Bureau}, {Burgasser}, {Burtin}, {Cano-D{\'\i}az},
  {Capasso}, {Cappellari}, {Carrera}, {Chabanier}, {Chaplin}, {Chapman},
  {Cherinka}, {Chiappini}, {Doohyun Choi}, {Chojnowski}, {Chung}, {Clerc},
  {Coffey}, {Comerford}, {Comparat}, {da Costa}, {Cousinou}, {Covey}, {Crane},
  {Cunha}, {da Silva Ilha}, {Dai}, {Damsted}, {Darling}, {Davidson}, {Davies},
  {Dawson}, {De}, {de la Macorra}, {De Lee}, {de Andrade Queiroz}, {Deconto
  Machado}, {de la Torre}, {Dell'Agli}, {du Mas des Bourboux},
  {Diamond-Stanic}, {Dillon}, {Donor}, {Drory}, {Duckworth}, {Dwelly},
  {Ebelke}, {Eftekharzadeh}, {Eigenbrot}, {Elsworth}, {Eracleous},
  {Erfanianfar}, {Escoffier}, {Fan}, {Farr}, {Fern{\'a}ndez-Trincado},
  {Feuillet}, {Finoguenov}, {Fofie}, {Fraser-McKelvie}, {Frinchaboy},
  {Fromenteau}, {Fu}, {Galbany}, {Garcia}, {Garc{\'\i}a-Hern{\'a}ndez}, {Garma
  Oehmichen}, {Ge}, {Geimba Maia}, {Geisler}, {Gelfand}, {Goddy},
  {Gonzalez-Perez}, {Grabowski}, {Green}, {Grier}, {Guo}, {Guy}, {Harding},
  {Hasselquist}, {Hawken}, {Hayes}, {Hearty}, {Hekker}, {Hogg}, {Holtzman},
  {Horta}, {Hou}, {Hsieh}, {Huber}, {Hunt}, {Ider Chitham}, {Imig}, {Jaber},
  {Jimenez Angel}, {Johnson}, {Jones}, {J{\"o}nsson}, {Jullo}, {Kim},
  {Kinemuchi}, {Kirkpatrick}, {Kite}, {Klaene}, {Kneib}, {Kollmeier}, {Kong},
  {Kounkel}, {Krishnarao}, {Lacerna}, {Lan}, {Lane}, {Law}, {Le Goff}, {Leung},
  {Lewis}, {Li}, {Lian}, {Lin}, {Long}, {Longa-Pe{\~n}a}, {Lundgren}, {Lyke},
  {Ted Mackereth}, {MacLeod}, {Majewski}, {Manchado}, {Maraston}, {Martini},
  {Masseron}, {Masters}, {Mathur}, {McDermid}, {Merloni}, {Merrifield},
  {M{\'e}sz{\'a}ros}, {Miglio}, {Minniti}, {Minsley}, {Miyaji}, {Mohammad},
  {Mosser}, {Mueller}, {Muna}, {Mu{\~n}oz-Guti{\'e}rrez}, {Myers}, {Nadathur},
  {Nair}, {Nandra}, {do Nascimento}, {Nevin}, {Newman}, {Nidever}, {Nitschelm},
  {Noterdaeme}, {O'Connell}, {Olmstead}, {Oravetz}, {Oravetz}, {Osorio},
  {Pace}, {Padilla}, {Palanque-Delabrouille}, {Palicio}, {Pan}, {Pan},
  {Parker}, {Paviot}, {Peirani}, {Pe{\~n}a Ram{\'r}ez}, {Penny}, {Percival},
  {Perez-Fournon}, {P{\'e}rez-R{\`a}fols}, {Petitjean}, {Pieri},
  {Pinsonneault}, {Poovelil}, {Povick}, {Prakash}, {Price-Whelan}, {Raddick},
  {Raichoor}, {Ray}, {Rembold}, {Rezaie}, {Riffel}, {Riffel}, {Rix}, {Robin},
  {Roman-Lopes}, {Rom{\'a}n-Z{\'u}{\~n}iga}, {Rose}, {Ross}, {Rossi},
  {Rowlands}, {Rubin}, {Salvato}, {S{\'a}nchez}, {S{\'a}nchez-Menguiano},
  {S{\'a}nchez-Gallego}, {Sayres}, {Schaefer}, {Schiavon}, {Schimoia},
  {Schlafly}, {Schlegel}, {Schneider}, {Schultheis}, {Schwope}, {Seo},
  {Serenelli}, {Shafieloo}, {Shamsi}, {Shao}, {Shen}, {Shetrone}, {Shirley},
  {Silva Aguirre}, {Simon}, {Skrutskie}, {Slosar}, {Smethurst}, {Sobeck},
  {Sodi}, {Souto}, {Stark}, {Stassun}, {Steinmetz}, {Stello}, {Stermer},
  {Storchi-Bergmann}, {Streblyanska}, {Stringfellow}, {Stutz}, {Su{\'a}rez},
  {Sun}, {Taghizadeh-Popp}, {Talbot}, {Tayar}, {Thakar}, {Theriault}, {Thomas},
  {Thomas}, {Tinker}, {Tojeiro}, {Toledo}, {Tremonti}, {Troup}, {Tuttle},
  {Unda-Sanzana}, {Valentini}, {Vargas-Gonz{\'a}lez}, {Vargas-Maga{\~n}a},
  {V{\'a}zquez-Mata}, {Vivek}, {Wake}, {Wang}, {Weaver}, {Weijmans}, {Wild},
  {Wilson}, {Wilson}, {Wolthuis}, {Wood-Vasey}, {Yan}, {Yang}, {Y{\`e}che},
  {Zamora}, {Zarrouk}, {Zasowski}, {Zhang}, {Zhao}, {Zhao}, {Zheng}, {Zheng},
  {Zhu}, \& {Zou}}]{ahumada20}
{Ahumada}, R., {Allende Prieto}, C., {Almeida}, A., {et~al.} 2020, \apjs, 249,
  3

\bibitem[{{Althaus} {et~al.}(2013){Althaus}, {Miller Bertolami}, \&
  {C{\'o}rsico}}]{althaus13}
{Althaus}, L.~G., {Miller Bertolami}, M.~M., \& {C{\'o}rsico}, A.~H. 2013,
  \aap, 557, A19

\bibitem[{{Amaro-Seoane} {et~al.}(2012){Amaro-Seoane}, {Aoudia}, {Babak},
  {Bin{\'e}truy}, {Berti}, {Boh{\'e}}, {Caprini}, {Colpi}, {Cornish},
  {Danzmann}, {Dufaux}, {Gair}, {Jennrich}, {Jetzer}, {Klein}, {Lang}, {Lobo},
  {Littenberg}, {McWilliams}, {Nelemans}, {Petiteau}, {Porter}, {Schutz},
  {Sesana}, {Stebbins}, {Sumner}, {Vallisneri}, {Vitale}, {Volonteri}, \&
  {Ward}}]{lisa12}
{Amaro-Seoane}, P., {Aoudia}, S., {Babak}, S., {et~al.} 2012, Classical and
  Quantum Gravity, 29, 124016

\bibitem[{{Bailer-Jones} {et~al.}(2021){Bailer-Jones}, {Rybizki}, {Fouesneau},
  {Demleitner}, \& {Andrae}}]{bailer21}
{Bailer-Jones}, C.~A.~L., {Rybizki}, J., {Fouesneau}, M., {Demleitner}, M., \&
  {Andrae}, R. 2021, \aj, 161, 147

\bibitem[{{B{\'e}dard} {et~al.}(2020){B{\'e}dard}, {Bergeron}, {Brassard}, \&
  {Fontaine}}]{bedard20}
{B{\'e}dard}, A., {Bergeron}, P., {Brassard}, P., \& {Fontaine}, G. 2020, \apj,
  901, 93

\bibitem[{{B{\'e}dard} {et~al.}(2017){B{\'e}dard}, {Bergeron}, \&
  {Fontaine}}]{bedard17}
{B{\'e}dard}, A., {Bergeron}, P., \& {Fontaine}, G. 2017, \apj, 848, 11

\bibitem[{{Bellm} {et~al.}(2019){Bellm}, {Kulkarni}, {Graham}, {Dekany},
  {Smith}, {Riddle}, {Masci}, {Helou}, {Prince}, {Adams}, {Barbarino},
  {Barlow}, {Bauer}, {Beck}, {Belicki}, {Biswas}, {Blagorodnova}, {Bodewits},
  {Bolin}, {Brinnel}, {Brooke}, {Bue}, {Bulla}, {Burruss}, {Cenko}, {Chang},
  {Connolly}, {Coughlin}, {Cromer}, {Cunningham}, {De}, {Delacroix}, {Desai},
  {Duev}, {Eadie}, {Farnham}, {Feeney}, {Feindt}, {Flynn}, {Franckowiak},
  {Frederick}, {Fremling}, {Gal-Yam}, {Gezari}, {Giomi}, {Goldstein},
  {Golkhou}, {Goobar}, {Groom}, {Hacopians}, {Hale}, {Henning}, {Ho}, {Hover},
  {Howell}, {Hung}, {Huppenkothen}, {Imel}, {Ip}, {Ivezi{\'c}}, {Jackson},
  {Jones}, {Juric}, {Kasliwal}, {Kaspi}, {Kaye}, {Kelley}, {Kowalski},
  {Kramer}, {Kupfer}, {Landry}, {Laher}, {Lee}, {Lin}, {Lin}, {Lunnan},
  {Giomi}, {Mahabal}, {Mao}, {Miller}, {Monkewitz}, {Murphy}, {Ngeow},
  {Nordin}, {Nugent}, {Ofek}, {Patterson}, {Penprase}, {Porter}, {Rauch},
  {Rebbapragada}, {Reiley}, {Rigault}, {Rodriguez}, {van Roestel}, {Rusholme},
  {van Santen}, {Schulze}, {Shupe}, {Singer}, {Soumagnac}, {Stein}, {Surace},
  {Sollerman}, {Szkody}, {Taddia}, {Terek}, {Van Sistine}, {van Velzen},
  {Vestrand}, {Walters}, {Ward}, {Ye}, {Yu}, {Yan}, \& {Zolkower}}]{ztf}
{Bellm}, E.~C., {Kulkarni}, S.~R., {Graham}, M.~J., {et~al.} 2019, \pasp, 131,
  018002

\bibitem[{{Bours} {et~al.}(2015){Bours}, {Marsh}, {G{\"a}nsicke}, {Tauris},
  {Istrate}, {Badenes}, {Dhillon}, {Gal-Yam}, {Hermes}, {Kengkriangkrai},
  {Kilic}, {Koester}, {Mullally}, {Prasert}, {Steeghs}, {Thompson}, \&
  {Thorstensen}}]{bours15}
{Bours}, M.~C.~P., {Marsh}, T.~R., {G{\"a}nsicke}, B.~T., {et~al.} 2015,
  \mnras, 450, 3966

\bibitem[{{Brown} {et~al.}(2010){Brown}, {Kilic}, {Allende Prieto}, \&
  {Kenyon}}]{brown10}
{Brown}, W.~R., {Kilic}, M., {Allende Prieto}, C., \& {Kenyon}, S.~J. 2010,
  \apj, 723, 1072

\bibitem[{{Brown} {et~al.}(2020{\natexlab{a}}){Brown}, {Kilic}, {B{\'e}dard},
  {Kosakowski}, \& {Bergeron}}]{brown20a}
{Brown}, W.~R., {Kilic}, M., {B{\'e}dard}, A., {Kosakowski}, A., \& {Bergeron},
  P. 2020{\natexlab{a}}, \apjl, 892, L35

\bibitem[{{Brown} {et~al.}(2011){Brown}, {Kilic}, {Hermes}, {Allende Prieto},
  {Kenyon}, \& {Winget}}]{brown11}
{Brown}, W.~R., {Kilic}, M., {Hermes}, J.~J., {et~al.} 2011, \apjl, 737, L23

\bibitem[{{Brown} {et~al.}(2020{\natexlab{b}}){Brown}, {Kilic}, {Kosakowski},
  {Andrews}, {Heinke}, {Ag{\"u}eros}, {Camilo}, {Gianninas}, {Hermes}, \&
  {Kenyon}}]{brown20b}
{Brown}, W.~R., {Kilic}, M., {Kosakowski}, A., {et~al.} 2020{\natexlab{b}},
  \apj, 889, 49

\bibitem[{{Burdge} {et~al.}(2019{\natexlab{a}}){Burdge}, {Coughlin}, {Fuller},
  {Kupfer}, {Bellm}, {Bildsten}, {Graham}, {Kaplan}, {Roestel}, {Dekany},
  {Duev}, {Feeney}, {Giomi}, {Helou}, {Kaye}, {Laher}, {Mahabal}, {Masci},
  {Riddle}, {Shupe}, {Soumagnac}, {Smith}, {Szkody}, {Walters}, {Kulkarni}, \&
  {Prince}}]{burdge19a}
{Burdge}, K.~B., {Coughlin}, M.~W., {Fuller}, J., {et~al.} 2019{\natexlab{a}},
  \nat, 571, 528

\bibitem[{{Burdge} {et~al.}(2019{\natexlab{b}}){Burdge}, {Fuller}, {Phinney},
  {van Roestel}, {Claret}, {Cukanovaite}, {Gentile Fusillo}, {Coughlin},
  {Kaplan}, {Kupfer}, {Tremblay}, {Dekany}, {Duev}, {Feeney}, {Riddle},
  {Kulkarni}, \& {Prince}}]{burdge19b}
{Burdge}, K.~B., {Fuller}, J., {Phinney}, E.~S., {et~al.} 2019{\natexlab{b}},
  \apjl, 886, L12

\bibitem[{{Burdge} {et~al.}(2020{\natexlab{a}}){Burdge}, {Prince}, {Fuller},
  {Kaplan}, {Marsh}, {Tremblay}, {Zhuang}, {Bellm}, {Caiazzo}, {Coughlin},
  {Dhillon}, {Gaensicke}, {Rodr{\'\i}guez-Gil}, {Graham}, {Hermes}, {Kupfer},
  {Littlefair}, {Mr{\'o}z}, {Phinney}, {van Roestel}, {Yao}, {Dekany}, {Drake},
  {Duev}, {Hale}, {Feeney}, {Helou}, {Kaye}, {Mahabal}, {Masci}, {Riddle},
  {Smith}, {Soumagnac}, \& {Kulkarni}}]{burdge20a}
{Burdge}, K.~B., {Prince}, T.~A., {Fuller}, J., {et~al.} 2020{\natexlab{a}},
  \apj, 905, 32

\bibitem[{{Burdge} {et~al.}(2020{\natexlab{b}}){Burdge}, {Coughlin}, {Fuller},
  {Kaplan}, {Kulkarni}, {Marsh}, {Bellm}, {Dekany}, {Duev}, {Graham},
  {Mahabal}, {Masci}, {Laher}, {Riddle}, {Soumagnac}, \& {Prince}}]{burdge20b}
{Burdge}, K.~B., {Coughlin}, M.~W., {Fuller}, J., {et~al.} 2020{\natexlab{b}},
  \apjl, 905, L7

\bibitem[{{Fuller} \& {Lai}(2013)}]{fuller13}
{Fuller}, J., \& {Lai}, D. 2013, \mnras, 430, 274

\bibitem[{{Groot} {et~al.}(2019){Groot}, {Bloemen}, \& {Jonker}}]{groot19}
{Groot}, P., {Bloemen}, S., \& {Jonker}, P. 2019, in The La Silla Observatory -
  From the Inauguration to the Future, 33

\bibitem[{{Hermes} {et~al.}(2012){Hermes}, {Kilic}, {Brown}, {Winget}, {Allende
  Prieto}, {Gianninas}, {Mukadam}, {Cabrera-Lavers}, \& {Kenyon}}]{hermes12}
{Hermes}, J.~J., {Kilic}, M., {Brown}, W.~R., {et~al.} 2012, \apjl, 757, L21

\bibitem[{{Istrate} {et~al.}(2016){Istrate}, {Marchant}, {Tauris}, {Langer},
  {Stancliffe}, \& {Grassitelli}}]{istrate16}
{Istrate}, A.~G., {Marchant}, P., {Tauris}, T.~M., {et~al.} 2016, \aap, 595,
  A35

\bibitem[{{Kilic} {et~al.}(2020){Kilic}, {B{\'e}dard}, {Bergeron}, \&
  {Kosakowski}}]{kilic20b}
{Kilic}, M., {B{\'e}dard}, A., {Bergeron}, P., \& {Kosakowski}, A. 2020,
  \mnras, 493, 2805

\bibitem[{{Kilic} {et~al.}(2010){Kilic}, {Brown}, {Allende Prieto}, {Kenyon},
  \& {Panei}}]{kilic10a}
{Kilic}, M., {Brown}, W.~R., {Allende Prieto}, C., {Kenyon}, S.~J., \& {Panei},
  J.~A. 2010, \apj, 716, 122

\bibitem[{{Kilic} {et~al.}(2014){Kilic}, {Brown}, {Gianninas}, {Hermes},
  {Allende Prieto}, \& {Kenyon}}]{kilic14}
{Kilic}, M., {Brown}, W.~R., {Gianninas}, A., {et~al.} 2014, \mnras, 444, L1

\bibitem[{{Korol} {et~al.}(2017){Korol}, {Rossi}, {Groot}, {Nelemans},
  {Toonen}, \& {Brown}}]{korol17}
{Korol}, V., {Rossi}, E.~M., {Groot}, P.~J., {et~al.} 2017, \mnras, 470, 1894

\bibitem[{{Kosakowski} {et~al.}(2020){Kosakowski}, {Kilic}, {Brown}, \&
  {Gianninas}}]{kosakowski20}
{Kosakowski}, A., {Kilic}, M., {Brown}, W.~R., \& {Gianninas}, A. 2020, \apj,
  894, 53

\bibitem[{{Lenz} \& {Breger}(2014)}]{period04}
{Lenz}, P., \& {Breger}, M. 2014, {Period04: Statistical analysis of large
  astronomical time series}, , , ascl:1407.009

\bibitem[{{Littenberg} \& {Cornish}(2019)}]{littenberg19}
{Littenberg}, T.~B., \& {Cornish}, N.~J. 2019, \apjl, 881, L43

\bibitem[{{Martin} {et~al.}(2005){Martin}, {Fanson}, {Schiminovich},
  {Morrissey}, {Friedman}, {Barlow}, {Conrow}, {Grange}, {Jelinsky},
  {Milliard}, {Siegmund}, {Bianchi}, {Byun}, {Donas}, {Forster}, {Heckman},
  {Lee}, {Madore}, {Malina}, {Neff}, {Rich}, {Small}, {Surber}, {Szalay},
  {Welsh}, \& {Wyder}}]{martin05}
{Martin}, D.~C., {Fanson}, J., {Schiminovich}, D., {et~al.} 2005, \apjl, 619,
  L1

\bibitem[{{McMahon} {et~al.}(2021){McMahon}, {R.}, {Banerji}, {M.}, {Gonzalez},
  {E.}, {Koposov}, {S.}, {Bejar}, {v.}, {Lodieu}, {N.}, {Rebolo}, {R.}, \& {The
  Vhs Collaboration}}]{mcmahon21}
{McMahon}, {R.}, G., {Banerji}, {et~al.} 2021, VizieR Online Data Catalog,
  II/367

\bibitem[{{Morris} \& {Naftilan}(1993)}]{morris93}
{Morris}, S.~L., \& {Naftilan}, S.~A. 1993, \apj, 419, 344

\bibitem[{{Mukadam} {et~al.}(2011){Mukadam}, {Owen}, {Mannery}, {MacDonald},
  {Williams}, {Stauffer}, \& {Miller}}]{mukadam11}
{Mukadam}, A.~S., {Owen}, R., {Mannery}, E., {et~al.} 2011, \pasp, 123, 1423

\bibitem[{{Nelemans} {et~al.}(2001){Nelemans}, {Yungelson}, \& {Portegies
  Zwart}}]{nelemans01b}
{Nelemans}, G., {Yungelson}, L.~R., \& {Portegies Zwart}, S.~F. 2001, \aap,
  375, 890

\bibitem[{{Nissanke} {et~al.}(2012){Nissanke}, {Vallisneri}, {Nelemans}, \&
  {Prince}}]{nissanke12}
{Nissanke}, S., {Vallisneri}, M., {Nelemans}, G., \& {Prince}, T.~A. 2012,
  \apj, 758, 131

\bibitem[{{Onken} {et~al.}(2019){Onken}, {Wolf}, {Bessell}, {Chang}, {Da
  Costa}, {Luvaul}, {Mackey}, {Schmidt}, \& {Shao}}]{skymapper19}
{Onken}, C.~A., {Wolf}, C., {Bessell}, M.~S., {et~al.} 2019, \pasa, 36, e033

\bibitem[{{Piro}(2019)}]{piro19}
{Piro}, A.~L. 2019, \apjl, 885, L2

\bibitem[{{Robson} {et~al.}(2019){Robson}, {Cornish}, \& {Liu}}]{robson19}
{Robson}, T., {Cornish}, N.~J., \& {Liu}, C. 2019, Classical and Quantum
  Gravity, 36, 105011

\bibitem[{{Ruiter} {et~al.}(2010){Ruiter}, {Belczynski}, {Benacquista},
  {Larson}, \& {Williams}}]{ruiter10}
{Ruiter}, A.~J., {Belczynski}, K., {Benacquista}, M., {Larson}, S.~L., \&
  {Williams}, G. 2010, \apj, 717, 1006

\bibitem[{{Shah} \& {Nelemans}(2014)}]{shah14}
{Shah}, S., \& {Nelemans}, G. 2014, \apj, 791, 76

\bibitem[{{Shah} {et~al.}(2013){Shah}, {Nelemans}, \& {van der Sluys}}]{shah13}
{Shah}, S., {Nelemans}, G., \& {van der Sluys}, M. 2013, \aap, 553, A82

\bibitem[{{Shah} {et~al.}(2012){Shah}, {van der Sluys}, \& {Nelemans}}]{shah12}
{Shah}, S., {van der Sluys}, M., \& {Nelemans}, G. 2012, \aap, 544, A153

\bibitem[{{Shporer} {et~al.}(2010){Shporer}, {Kaplan}, {Steinfadt}, {Bildsten},
  {Howell}, \& {Mazeh}}]{shporer10}
{Shporer}, A., {Kaplan}, D.~L., {Steinfadt}, J. D.~R., {et~al.} 2010, \apjl,
  725, L200

\bibitem[{{Zucker} {et~al.}(2007){Zucker}, {Mazeh}, \& {Alexander}}]{zucker07}
{Zucker}, S., {Mazeh}, T., \& {Alexander}, T. 2007, \apj, 670, 1326

\end{thebibliography}

\end{document}